\documentclass[letterpaper,10pt]{article}

\usepackage{style/osameet3} 
\usepackage[utf8]{inputenc}
\usepackage{glossaries}
\usepackage[USenglish]{babel}

\newcommand{\SetCapsType}{normalcaps}

\makeatletter
\providecommand{\SetCapsType}{smallcaps}

\long\def\@scTrue{smallcaps}
\long\def\@scFalse{normalcaps}
\newcommand{\acroSCaps}[1]{%
 \begingroup
  \ifx\SetCapsType\@scTrue 
    \textsc{#1}%
  \else
    \MakeUppercase{#1}%
  \fi
  \endgroup
}
\makeatother

\newcommand{\nAcronym}[4][]{%
	\newacronym[#1]{#2}{#3}{#4}
}

\makeatletter
\@ifpackageloaded{babel}{%
    \newcommand{\usuk}[2]{%
        \iflanguage{USenglish}{#1}{#2}%
    }%
}{%
    \newcommand{\usuk}[2]{%
        #1%
    }%
}%
\makeatother

\usepackage[shortcuts]{extdash}
\newcommand{\qam}[1]{
    \ifglsused{QAM}%
        {#1\=/\gls{QAM}}%
        {#1\=/ary \gls{QAM}%
    }%
}%

\nAcronym{2A8PSK}{\acroSCaps{2a8psk}}{2-ary amplitude 8-ary phaseshift keying}

\nAcronym{3CCMCF}{\acroSCaps{3cc-mcf}}{3-core coupled-core multi-core fiber}

\nAcronym{4D}{\acroSCaps{4d}}{four-dimensional}
\nAcronym{4D64PRS}{\acroSCaps{4d-64prs}}{\usuk{four-dimensional 64-ary polarization-ring-switching}{four-dimensional 64-ary polarisation-ring-switching}}

\nAcronym{5B4D2A8PSK}{\acroSCaps{5b4d-2a8psk}}{5-bit four-dimensional two-amplitude 8-ary phase-shift keying}

\nAcronym{6B4D2A8PSK}{\acroSCaps{6b4d-2a8psk}}{6-bit four-dimensional two-amplitude 8-ary phase-shift keying}

\nAcronym{8D}{\acroSCaps{8d}}{eight-dimensional}\nAcronym{8D2048PRS}{\acroSCaps{8d-2048prs}}{eight-dimensional 2048-ary polarization-ring-switching}
\nAcronym{8D2048PRST1}{\acroSCaps{8d-2048prs-t1}}{eight-dimensional 2048-ary polarization-ring-switching type 1}
\nAcronym{8D2048PRST2}{\acroSCaps{8d-2048prs-t2}}{eight-dimensional 2048-ary polarization-ring-switching type 2}

\nAcronym{ABC}{\acroSCaps{abc}}{automatic bias controller}
\nAcronym{ADC}{\acroSCaps{adc}}{analog-to-digital converter}
\nAcronym{AIR}{\acroSCaps{air}}{achievable information rate}
\nAcronym{AOM}{\acroSCaps{aom}}{acoustic optical modulator}
\nAcronym{API}{\acroSCaps{api}}{application programming interface}
\nAcronym{AR}{\acroSCaps{ar}}{achievable rate}
\nAcronym{ASE}{\acroSCaps{ase}}{amplified spontaneous emission}
\nAcronym{ASK}{\acroSCaps{ask}}{amplitude-shift keying}
\nAcronym{ASIC}{\acroSCaps{asic}}{application-specific integrated circuit}
\nAcronym{ARRWG}{\acroSCaps{a}rr\acroSCaps{wg}}{arrayed-waveguide grating}
\nAcronym{AWG}{\acroSCaps{awg}}{arbitrary-waveform generator}
\nAcronym{AWGN}{\acroSCaps{awgn}}{additive white Gaussian noise}

\nAcronym{BCH}{\acroSCaps{bch}}{Bose-Chaudhuri-Hocquenghem}
\nAcronym{BER}{\acroSCaps{ber}}{bit error rate}
\nAcronym{BICM}{\acroSCaps{bicm}}{bit-interleaved coded modulation}
\nAcronym{BMD}{\acroSCaps{bmd}}{bit-metric decoding}
\nAcronym{BPD}{\acroSCaps{bpd}}{balanced photo-diode}
\nAcronym{BPF}{\acroSCaps{bpf}}{bandpass filter}
\nAcronym{BPS}{\acroSCaps{bps}}{blind phase search}
\nAcronym{BPSK}{\acroSCaps{bpsk}}{binary phase-shift keying}
\nAcronym{BRGC}{\acroSCaps{brgc}}{binary reflected Gray code}
\nAcronym{BTB}{\acroSCaps{btb}}{back-to-back}

\nAcronym{CAGR}{\acroSCaps{cagr}}{compound annual growth rate}
\nAcronym{CCDM}{\acroSCaps{ccdm}}{constant composition distribution matching}
\nAcronym{CCF}{\acroSCaps{ccf}}{\usuk{coupled-core fiber}{coupled-core fibre}}
\nAcronym{CD}{\acroSCaps{cd}}{chromatic dispersion}
\nAcronym{CIR}{\acroSCaps{cir}}{channel impulse response}
\nAcronym{CMA}{\acroSCaps{cma}}{constant modulus algorithm}
\nAcronym{CMUX}{\acroSCaps{cmux}}{core multiplexer}
\nAcronym{ChUT}{\acroSCaps{chut}}{channel under test}
\nAcronym{CUT}{\acroSCaps{cut}}{channel under test}
\nAcronym{CPE}{\acroSCaps{cpe}}{carrier phase estimation}
\nAcronym{CPU}{\acroSCaps{cpu}}{central processing unit}
\nAcronym{CSPR}{\acroSCaps{cspr}}{carrier-to-signal power ratio}
\nAcronym{CW}{\acroSCaps{cw}}{continuous wave}

\nAcronym{DA}{\acroSCaps{da}}{driver amplifier}
\nAcronym{DAC}{\acroSCaps{dac}}{digital-to-analog converter}
\nAcronym{DCF}{\acroSCaps{dcf}}{\usuk{dispersion compensated fiber}{dispersion compensated fibre}}
\nAcronym{DCI}{\acroSCaps{dci}}{data center interconnect}
\nAcronym{DDLMS}{\acroSCaps{dd-lms}}{decision-directed least mean square}
\nAcronym{DFB}{\acroSCaps{dfb}}{distributed feedback}
\nAcronym{DGD}{\acroSCaps{dgd}}{differential group delay}
\nAcronym{DH}{\acroSCaps{dh}}{digital holography}
\nAcronym{DM}{\acroSCaps{dm}}{distribution matcher}
\nAcronym{DMA}{\acroSCaps{dma}}{direct memory access}
\nAcronym{DMD}{\acroSCaps{dmd}}{differential mode delay}
\nAcronym{DMG}{\acroSCaps{dmg}}{differential modal gain}
\nAcronym{DMGD}{\acroSCaps{dmgd}}{differential mode group delay}
\nAcronym{DML}{\acroSCaps{dml}}{directly-modulated laser}
\nAcronym{DP}{\acroSCaps{dp}}{\usuk{dual-polarization}{dual-polarisation}}
\nAcronym{DPC}{\acroSCaps{dpc}}{digital pre-compensation}
\nAcronym{DPE}{\acroSCaps{dpe}}{digital pre-emphasis}
\nAcronym{DPIQ}{\acroSCaps{dp-iqm}}{\usuk{dual-polarization IQ-modulator}{dual-polarisation IQ-modulator}}
\nAcronym{DPLL}{\acroSCaps{dpll}}{digital phase-locked loop}
\nAcronym{DQPSK}{\acroSCaps{dqpsk}}{differential quaternary phase-shift-keying}
\nAcronym{DRA}{\acroSCaps{dra}}{distributed Raman amplifier}
\nAcronym{DRE}{\acroSCaps{dre}}{digital resolution enhancer}
\nAcronym{DSF}{\acroSCaps{dsf}}{\usuk{dispersion-shifted fiber}{dispersion-shifted fibre}}
\nAcronym{DSO}{\acroSCaps{dso}}{digital sampling oscilloscope}
\nAcronym{DSP}{\acroSCaps{dsp}}{digital signal processing}
\nAcronym{DUT}{\acroSCaps{dut}}{device-under-test}
\nAcronym{DWDM}{\acroSCaps{dwdm}}{dense wavelength-division multiplexing}

\nAcronym{EAM}{\acroSCaps{eam}}{electro-absorption modulator}
\nAcronym{ECL}{\acroSCaps{ecl}}{external cavity laser}
\nAcronym{ED}{\acroSCaps{ed}}{Eucledian distance}
\nAcronym{EDFA}{\acroSCaps{edfa}}{\usuk{erbium-doped fiber amplifier}{erbium-doped fibre amplifier}}
\nAcronym{ENOB}{\acroSCaps{enob}}{effective number of bits}
\nAcronym{ESS}{\acroSCaps{ess}}{enumerative sphere shaping}

\nAcronym{FD}{\acroSCaps{fd}}{frequency domain}
\nAcronym{FDE}{\acroSCaps{fde}}{\usuk{frequency domain equalizer}{frequency domain equaliser}}
\nAcronym{FEC}{\acroSCaps{fec}}{forward error correction}
\nAcronym{FFT}{\acroSCaps{fft}}{fast Fourier transform}
\nAcronym{FIR}{\acroSCaps{fir}}{finite impulse response}
\nAcronym{FMF}{\acroSCaps{fmf}}{\usuk{few-mode fiber}{few-mode fibre}}
\nAcronym[plural=FM-MCF, firstplural=\usuk{few-mode multi-core fibers}{few-mode multi-core fibres}]{FM-MCF}{\acroSCaps{fm-mcf}}{\usuk{few-mode multi-core fiber}{few-mode multi-core fibre}}
\nAcronym{FPGA}{\acroSCaps{fpga}}{field-programmable gate array}
\nAcronym{FWM}{\acroSCaps{fwm}}{four-wave mixing}

\nAcronym{GD}{\acroSCaps{gd}}{group delay}
\nAcronym{GI}{\acroSCaps{gi}}{graded-index}
\nAcronym{GFF}{\acroSCaps{gff}}{gain flattening filter}
\nAcronym{GIMMF}{\acroSCaps{gi-mmf}}{\usuk{graded-index multi-mode fiber}{graded-index multi-mode fibre}}
\nAcronym{GMI}{\acroSCaps{gmi}}{\usuk{generalized mutual information}{generalised mutual information}}
\nAcronym{GNSE}{\acroSCaps{gnse}}{generalized nonlinear Schr\"{o}dinger equation}
\nAcronym{GPU}{\acroSCaps{gpu}}{graphics processing unit}
\nAcronym{GS}{\acroSCaps{gs}}{geometric shaping}
\nAcronym{GV}{\acroSCaps{gv}}{group-velocity}
\nAcronym{GVD}{\acroSCaps{gvd}}{group-velocity dispersion}
\nAcronym{GPIO}{\acroSCaps{gpio}}{general purpose input output}

\nAcronym{HDFEC}{\acroSCaps{hd-fec}}{hard decision forward error correction}

\nAcronym[plural=IL, firstplural=insertion losses (\textsc{il})]{IL}{\acroSCaps{il}}{insertion loss}
\nAcronym{IFFT}{\acroSCaps{ifft}}{inverse fast Fourier transform}
\nAcronym{IMDD}{\acroSCaps{im/dd}}{intensity-modulation direct-detection}
\nAcronym{IQM}{\acroSCaps{iqm}}{in-phase and quadrature modulator}
\nAcronym{ISI}{\acroSCaps{isi}}{intersymbol interference}

\nAcronym{KK}{\acroSCaps{kk}}{Kramers-Kronig}

\nAcronym{LCOS}{\acroSCaps{LCoS}}{liquid crystal on silicon}
\nAcronym{LDPC}{\acroSCaps{ldpc}}{low-density parity-check}
\nAcronym{LEAF}{\acroSCaps{leaf}}{\usuk{large effective area fiber}{large effective area fibre}}
\nAcronym{LFSR}{\acroSCaps{lfsr}}{linear-feedback shift register}
\nAcronym{LMS}{\acroSCaps{lms}}{least means square}
\nAcronym{LLR}{\acroSCaps{llr}}{log-likelihood ratio}
\nAcronym{LO}{\acroSCaps{lo}}{local oscillator}
\nAcronym{LP}{\acroSCaps{lp}}{\usuk{linearly polarized}{linearly polarised}}
\nAcronym{LSPS}{\acroSCaps{lsps}}{\usuk{loop-synchronized polarization scrambler}{loop-synchronised polarisation scrambler}}
\nAcronym{LUT}{\acroSCaps{lut}}{lookup table}

\nAcronym{MB}{\acroSCaps{mb}}{Maxwell-Bolzmann}
\nAcronym{MCF}{\acroSCaps{mcf}}{\usuk{multi-core fiber}{multi-core fibre}}
\nAcronym{MDG}{\acroSCaps{mdg}}{mode dependent gain}
\nAcronym[plural=MDL, firstplural=mode-dependent losses (\textsc{mdl})]{MDL}{\acroSCaps{mdl}}{mode-dependent loss}
\nAcronym{MDM}{\acroSCaps{mdm}}{mode division multiplexing}
\nAcronym{MEMS}{\acroSCaps{mems}}{micro-electro-mechanical systems}
\nAcronym{MF}{\acroSCaps{mf}}{matched filter}
\nAcronym{MI}{\acroSCaps{mi}}{mutual information}
\nAcronym{MIMO}{\acroSCaps{mimo}}{multiple-input multiple-output}
\nAcronym{ML}{\acroSCaps{ml}}{machine learning}
\nAcronym{MMA}{\acroSCaps{mma}}{multi-modulus algorithm}
\nAcronym{MMF}{\acroSCaps{mmf}}{\usuk{multi-mode fiber}{multi-mode fibre}}
\nAcronym{MMSE}{\acroSCaps{mmse}}{minimum mean squared error}
\nAcronym{MP}{\acroSCaps{mp}}{minimum phase}
\nAcronym{MPLC}{\acroSCaps{mplc}}{multi-plane light converter}
\nAcronym{MSE}{\acroSCaps{mse}}{mean squared error}
\nAcronym{MUX}{\acroSCaps{mux}}{multiplexer}
\nAcronym{MZM}{\acroSCaps{mzm}}{Mach-Zehnder modulator}
\nAcronym{MZI}{\acroSCaps{mzi}}{Mach-Zehnder interferometer}

\nAcronym{NA}{\acroSCaps{na}}{numerical aperture}
\nAcronym{NF}{\acroSCaps{nf}}{noise figure}
\nAcronym{NGMI}{\acroSCaps{ngmi}}{\usuk{normalized generalized mutual information}{normalised generalised mutual information}}
\nAcronym{NLSE}{\acroSCaps{nlse}}{nonlinear Schr\"{o}ding equation}
\nAcronym{NIR}{\acroSCaps{nir}}{near-infrared}

\nAcronym{OBTB}{\acroSCaps{obtb}}{optical back-to-back}
\nAcronym{ODE}{\acroSCaps{ode}}{ordinary differential equation}
\nAcronym{ODL}{\acroSCaps{odl}}{optical delay line}
\nAcronym{OFC}{\acroSCaps{ofc}}{Optical Fiber Communications Conference}
\nAcronym{OFDR}{\acroSCaps{ofdr}}{optical frequency-domain reflectometer}
\nAcronym{OFDM}{\acroSCaps{ofdm}}{orthogonal frequency division multiplexing}
\nAcronym{OMFT}{\acroSCaps{omft}}{optical multi-format transmitter}
\nAcronym{OOK}{\acroSCaps{ook}}{on-off keying}
\nAcronym{OPLL}{\acroSCaps{opll}}{optical phase-locked loop}
\nAcronym{OSA}{\acroSCaps{osa}}{\usuk{optical spectrum analyzer}{optical spectrum analyser}}
\nAcronym{OSNR}{\acroSCaps{osnr}}{optical signal-to-noise ratio}
\nAcronym{OTDR}{\acroSCaps{otdr}}{optical time-domain reflectometer}
\nAcronym{OTF}{\acroSCaps{otf}}{optical tunable filter}
\nAcronym{OVNA}{\acroSCaps{ovna}}{\usuk{optical vector network analyzer}{optical vector network analyser}}

\nAcronym{PAM}{\acroSCaps{pam}}{pulse-amplitude modulation}
\nAcronym{PAS}{\acroSCaps{pas}}{probabilistic amplitude shaping}
\nAcronym{PAPR}{\acroSCaps{papr}}{peak-to-avarage power ratio}
\nAcronym{PBC}{\acroSCaps{pbc}}{\usuk{polarization beam combiner}{polarisation beam combiner}}
\nAcronym{PBS}{\acroSCaps{pbs}}{polarization beam splitter}
\nAcronym{PCVD}{\acroSCaps{pcvd}}{plasma chemical vapor depostion}
\nAcronym{PD}{\acroSCaps{pd}}{photodiode}
\nAcronym{PDL}{\acroSCaps{pdl}}{polarization-dependent loss}
\nAcronym{PDM}{\acroSCaps{pdm}}{\usuk{polarization-division multiplexing}{polarisation-division multiplexing}}
\nAcronym{PL}{\acroSCaps{pl}}{photonic lantern}
\nAcronym{PMBPSK}{\acroSCaps{pm-bpsk}}{polarization-multiplexed binary phase-shift-keying}
\nAcronym{PMQPSK}{\acroSCaps{pm-qpsk}}{polarization-multiplexed quaternary phase-shift-keying}
\nAcronym{PM8QAM}{\acroSCaps{pm-8qam}}{polarization-multiplexed 8-ary quadrature amplitude modulation}
\nAcronym{PMD}{\acroSCaps{pmd}}{polarization mode dispersion}
\nAcronym{PNOB}{\acroSCaps{pnob}}{physical number of bits}
\nAcronym{PON}{\acroSCaps{pon}}{passive-optical network}
\nAcronym{PRBS}{\acroSCaps{prbs}}{pseudorandom bit sequence}
\nAcronym{PS}{\acroSCaps{ps}}{probabilistic shaping}
\nAcronym{PSK}{\acroSCaps{psk}}{phase-shift keying}
\nAcronym{PSP}{\acroSCaps{psp}}{\usuk{principal states of polarization}{principal states of polarisation}}

\nAcronym{QAM}{\acroSCaps{qam}}{quadrature amplitude modulation}
\nAcronym{QPSK}{\acroSCaps{qpsk}}{quaternary phase-shift-keying}

\nAcronym{RAM}{\acroSCaps{ram}}{random-access memory}
\nAcronym{RF}{\acroSCaps{rf}}{radio frequency}
\nAcronym{RRC}{\acroSCaps{rrc}}{root-raised-cosine}
\nAcronym{ROADM}{\acroSCaps{roadm}}{reconfigurable optical add-drop multiplexer}
\nAcronym{ROI}{\acroSCaps{roi}}{region of interest}

\nAcronym{SA}{\acroSCaps{sa}}{simulated annealing}
\nAcronym{SamPerSym}{\acroSCaps{sps}}{samples per symbol}
\nAcronym{SBS}{\acroSCaps{sbs}}{stimulated Brillouin scattering}
\nAcronym{SDFEC}{\acroSCaps{sd-fec}}{soft decision forward error correction}
\nAcronym{SDM}{\acroSCaps{sdm}}{space-division multiplexing}
\nAcronym{SE}{\acroSCaps{se}}{spectral efficiency}
\nAcronym{SER}{\acroSCaps{ser}}{symbol error rate}
\nAcronym{SI}{\acroSCaps{si}}{step index}
\nAcronym{SLM}{\acroSCaps{slm}}{spatial light modulator}
\nAcronym{SMF}{\acroSCaps{smf}}{\usuk{single-mode fiber}{single-mode fibre}}
\nAcronym{SNR}{\acroSCaps{snr}}{signal-to-noise ratio}
\nAcronym{SOA}{\acroSCaps{soa}}{semiconductor optical amplifier}
\nAcronym[firstplural=\usuk{states of polarization (\acroSCaps{sop})}{states of polarisation (\acroSCaps{sop})}]{SOP}{\acroSCaps{sop}}{\usuk{state of polarization}{state of polarisation}}
\nAcronym{SPM}{\acroSCaps{spm}}{self-phase modulation}
\nAcronym{SPS}{\acroSCaps{sps}}{\usuk{synchronized polarization scrambler}{synchronised polarisation scrambler}}
\nAcronym{SRS}{\acroSCaps{srs}}{stimulated Raman scattering}
\nAcronym{SSBI}{\acroSCaps{ssbi}}{signal-signal beat interference}
\nAcronym{SSFM}{\acroSCaps{ssfm}}{split-step Fourier method}
\nAcronym{SSMF}{\acroSCaps{ssmf}}{\usuk{standard single-mode fiber}{standard single-mode fibre}}
\nAcronym{STL}{\acroSCaps{stl}}{swept tunable laser}
\nAcronym{SVD}{\acroSCaps{svd}}{singular value decomposition}
\nAcronym{SWI}{\acroSCaps{swi}}{swept wavelength interferometry}

\nAcronym{TD}{\acroSCaps{td}}{time domain}
\nAcronym{TDE}{\acroSCaps{tde}}{\usuk{time domain equalizer}{time domain equaliser}}
\nAcronym{TDMSDM}{\acroSCaps{tdm-sdm}}{time-domain multiplexed space-division multiplexing}
\nAcronym{TE}{\acroSCaps{te}}{transverse electric}
\nAcronym{TIA}{\acroSCaps{tia}}{trans-impedance amplifier}
\nAcronym{TM}{\acroSCaps{TM}}{transverse magnetic}
\nAcronym{TH4D}{\acroSCaps{th-4d}}{time domain hybrid four-dimensional}
\nAcronym{TH4D2A8PSK}{\acroSCaps{th-4d-2a8psk}}{time domain hybrid four-dimensional two-amplitude eight-phase shift keying}

\nAcronym{VCSEL}{\acroSCaps{vcsel}}{vertical-cavity surface emitting laser}
\nAcronym{VOA}{\acroSCaps{voa}}{variable optical attenuator}

\nAcronym{WDM}{\acroSCaps{wdm}}{wavelength-division multiplexing}
\nAcronym{WGA}{\acroSCaps{wga}}{weakly guiding approximation}
\nAcronym{WGN}{\acroSCaps{wgn}}{white Gaussian noise}
\nAcronym{WSS}{\acroSCaps{wss}}{wavelength selective switch}

\nAcronym{XPM}{\acroSCaps{xpm}}{cross-phase modulation}
\nAcronym{XT}{\acroSCaps{xt}}{cross-talk}

\nAcronym{3DWG}{\acroSCaps{3dwg}}{3D-waveguide} 

\usepackage{xcolor}
\usepackage{float}
\AtEndPreamble{\DeclareSymbolFont{operators}{OT1}{\rmdefault}{m}{n}}
\usepackage[per-mode=symbol]{siunitx}
\usepackage{lipsum}

\usepackage{silence}
\usepackage[subrefformat=parens]{subcaption}

\usepackage{amsmath,amssymb}
\usepackage[colorlinks=true,bookmarks=false,citecolor=blue,urlcolor=blue]{hyperref} 
\usepackage[capitalise]{cleveref}

\begin{document}

\title{10,000~km Straight-line Transmission using a Real-time Software-defined GPU-Based Receiver}

\author{Sjoerd~van~der~Heide\textsuperscript{1,2},
        Ruben~S.~Luis\textsuperscript{1},
        Benjamin~J.~Puttnam\textsuperscript{1},
        Georg~Rademacher\textsuperscript{1},
        Ton~Koonen\textsuperscript{2},
        Satoshi~Shinada\textsuperscript{1},
        Yohinari~Awaji\textsuperscript{1},
        Hideaki~Furukawa\textsuperscript{1},
        and Chigo~Okonkwo\textsuperscript{2}
        }

\address{
(1) National Institute of Information and Communication Technology, 4-2-1 Nukui-Kitamachi, Koganei, Japan\\
(2) High-Capacity Optical Transmission Laboratory, Eindhoven University of Technology, the Netherlands\\
}
\email{s.p.v.d.heide@tue.nl}

\vspace{-6mm}
\begin{abstract}
Real-time operation of a software-defined, GPU-based optical receiver is demonstrated over a 100-span straight-line optical link. Performance of minimum-phase \acrlong{KK} 4-, 8-, 16-, 32-, and 64-QAM signals are evaluated at various distances.
\end{abstract}

\begin{figure}[!b]
    \vspace{-5mm}
    \renewcommand\figurename{Fig.}
    \centering
    \includegraphics[width=\textwidth]{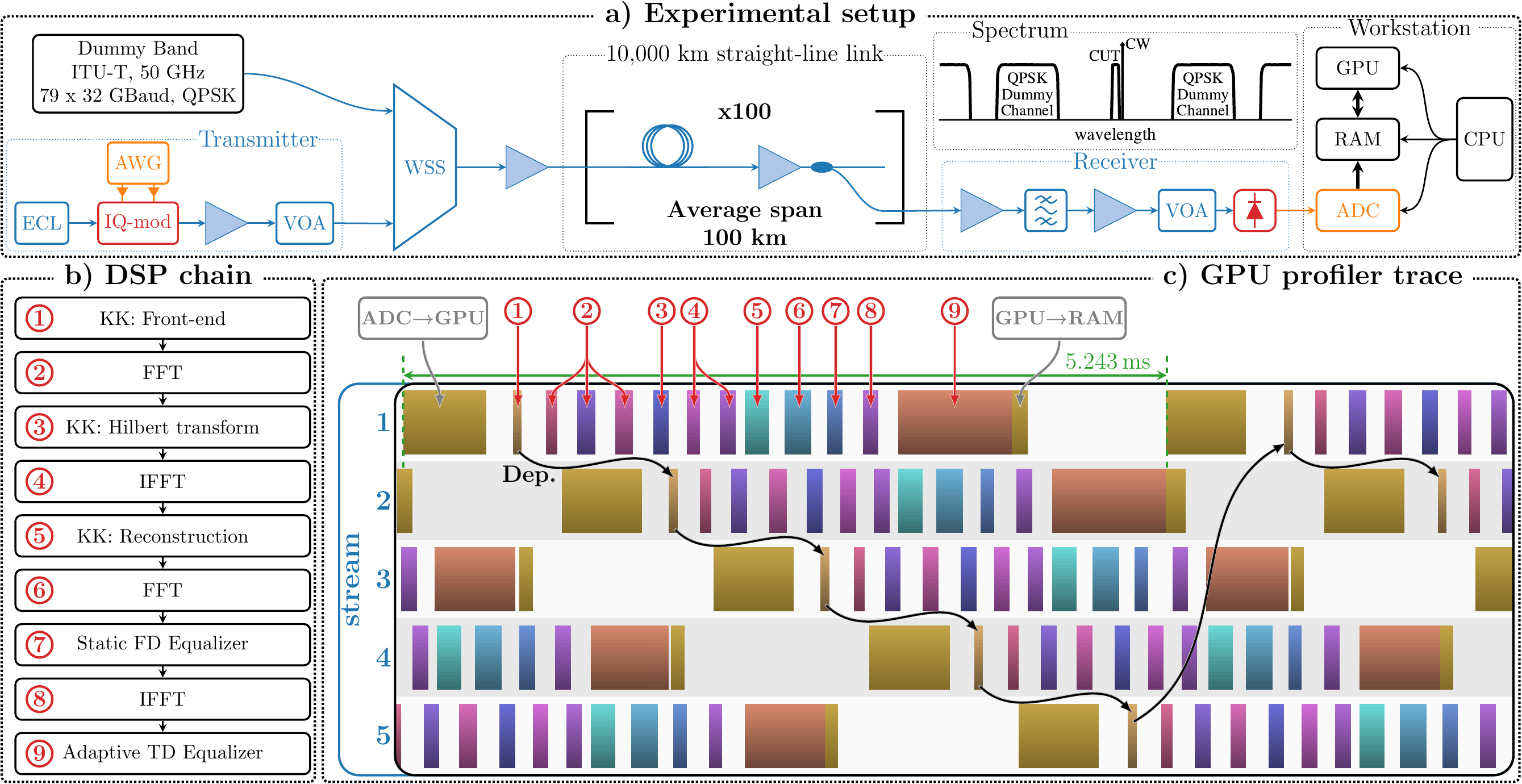}
    \caption{Experimental setup. a) transmission link; b) DSP chain and corresponding c) GPU profiler trace.}
    \label{fig:expsetup}
    \vspace{-5mm}
\end{figure}

\section{Introduction}

General purpose \glspl{GPU} for real-time digital signal processing have been recently shown as a potential alternative to the use of \glspl{FPGA} or \glspl{ASIC} \cite{Li_realtime_LDPC, suzuki_fec,suzuki_phy_2020,suzuki_real-time_2020,ECOC_realtimeGPU,gpuJLT}. The development of \glspl{GPU} has seen more than a decade of steady exponential improvements in computation capacity (45\% yearly increase \cite{winzer_scaling_2017}) and energy efficiency (25\% yearly increase \cite{sun_summarizing_2019}), which has enabled its use for real-time \gls{FEC} decoding \cite{Li_realtime_LDPC, suzuki_fec} and \gls{DQPSK} detection \cite{suzuki_real-time_2020}. Recently, it was experimentally demonstrated for flexible multi-modulation format detection using directly detected pulse-amplitude modulated signals, and coherently detected \gls{QAM} signals \cite{ECOC_realtimeGPU,gpuJLT} using \gls{KK} detection \cite{mecozzi_kramers_2016}.


In this work, we demonstrate the potential of real-time GPU-based \gls{DSP} for long distance transmission. We use a \SI{10000}{km} straight-line link consisting of 100 spans with an average span length of \SI{100}{km}, to transmit a \SI{1}{GBaud} \gls{MP} \gls{QAM} signal to be coherently detected and recovered in real-time by a GPU-based \gls{KK} coherent receiver. This test signal is combined with \SI{32}{GBaud} \qam{4} dummy channels to emulate full c-band transmission. The receiver includes equalization, which automatically handles dispersion compensation in the digital domain. Successful transmission reaching the 20\% overhead \gls{HDFEC} threshold \cite{agrell_information-theoretic_2018} is achieved after \SI{10000}{km} for \qam{4}, \SI{7600}{km} for \qam{8}, \SI{5600}{km} for \qam{16}, \SI{3600}{km} for \qam{32}, and \SI{1600}{km} for \qam{64}. For all cases, we optimized \gls{CSPR} and launch power. These results show the potential of the proposed approach for long distance transmission, particularly for applications relating to long distance datacenter-to-datacenter communications without the use of third party systems.

\newpage
\section{Experimental setup}
\cref{fig:expsetup}-a) shows the experimental setup. At the transmitter, the test signal was produced by modulating the light from a \SI{100}{kHz} linewidth \gls{ECL} operating at \SI{1550.51}{nm} with an \gls{IQM}. The modulator was driven by an \gls{AWG} operating at \SI{12}{GS/s} to produce an \gls{MP} signal for coherent \gls{KK} detection. The latter was digitally produced by combining a \SI{1}{GBaud} \gls{QAM} information sequence mapped onto a 1\% roll-off, \gls{RRC} signal and combined with a carrier tone at \SI{0.516}{GHz}. The test signal was combined with a fully loaded c-band \gls{WDM} signal using a \gls{WSS}. The dummy signal consisted of 79 carriers positioned according to the ITU-T \SI{50}{GHz} grid with one gap at \SI{1550.51}{nm} for the test signal. Each carrier was modulated with \SI{32}{GBaud} \qam{4} signals. After amplification by an \gls{EDFA}, the signal was transmitted through a 100~span straight-line link. The average span length was approximately \SI{100}{km} with monitoring taps at various distances.The transmission fiber was designed for submarine transmission and had an attenuation parameter of \SI{0.154}{dB/km} and effective area of \SI{112}{\micro\metre\squared}. The total launch power at the input of each span was \SI{20}{dBm} and \glspl{WSS} were placed every 20 spans to flatten the transmission spectrum. The average power of the test channel was set at a variable offset from the average power of the dummy channels and adjusted to optimize the trade-off between the impacts of fiber nonlinarities and noise. At the receiver, the \gls{WDM} signals tapped off the transmission link were first amplified and then filtered by a \SI{5}{GHz} \gls{BPF} to extract the test signal. The latter was then amplified again and detected using a \SI{6.5}{GHz} photodetector. The recovered electrical signal was then sent to a workstation, where it was digitized by a \SI{1}{GHz} bandwidth \SI{4}{GS/s} \gls{ADC}, and processed in real time on a \gls{GPU}.



\begin{figure}[b]
    \renewcommand\figurename{Fig.}
    \centering
    \begin{subfigure}{0.5\textwidth}
        \includegraphics[width=\textwidth]{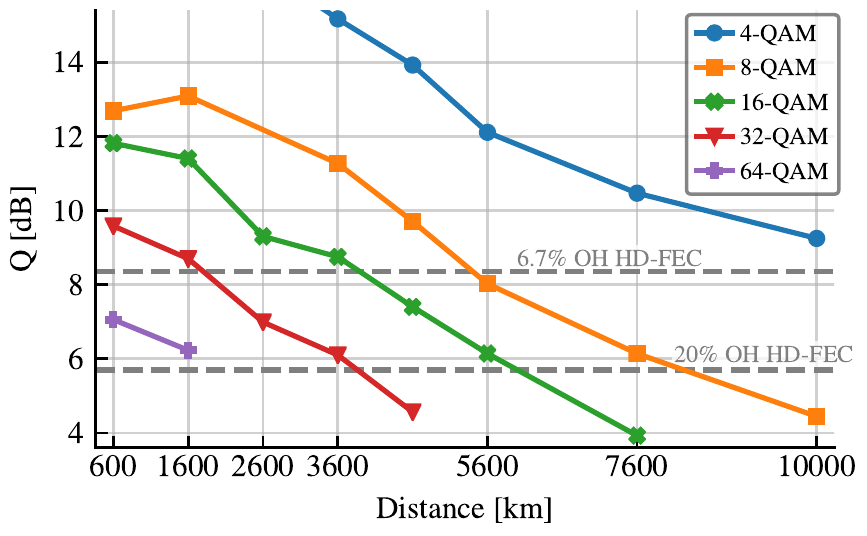}
        \vspace*{-5mm}\caption{Q-factor versus distance.}
        \label{fig:distance}
    \end{subfigure}%
    \begin{subfigure}{0.5\textwidth}
        \includegraphics[width=\textwidth]{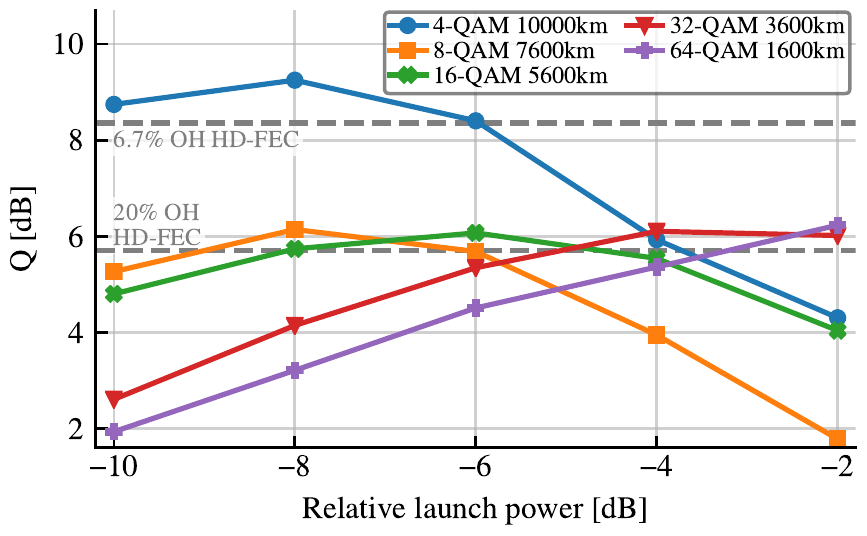}
        \vspace*{-5mm}\caption{Q-factor versus launch power.}
        \label{fig:launchpower}
    \end{subfigure}%
    \vspace{-5mm}
    \caption{Experimental results.}
    \label{fig:resultstop}
    \vspace{-5mm}
\end{figure}

\cref{fig:expsetup}-b) shows the real-time \gls{KK} signal processing chain. Its GPU-based implementation has been fully described in \cite{gpuJLT}. Buffers containing 2\textsuperscript{22} samples are transferred using \gls{DMA} from the \SI{4}{GS/s} \gls{ADC} to the \gls{GPU}. The first step of \gls{DSP} converts the samples to floating point and performs \gls{KK} front-end consisting of the square-root and logarithm operations. The phase of the optical signal, obtained through a Hilbert transform enabled by a pair of 1024-point 100\% overlap-save \glspl{FFT}\glsunset{IFFT}, is combined with the amplitude to digitally reconstruct the optical field \cite{mecozzi_kramers_2016}, which is subsequently downshifted to DC for further processing. Frequency-domain static equalization and downsampling from 4 to 2 samples-per-symbol is performed by multiplication with an offline-optimized filter enabled by another \gls{FFT} and \gls{IFFT} pair. Finally, the signal is further filtered by a four-tap adaptive time-domain \gls{DDLMS} widely-linear \cite{Silva_WidelyLinear} equalizer. The decisions made by the equalizer are demapped and sent to \gls{RAM}. Real-time operation of this process is illustrated by the GPU profiler trace in \cref{fig:expsetup}-c). It is shown that multiple processing streams work on different buffers in parallel whilst dependencies between buffers are handled through \textit{events} and marked by dark arrows in \cref{fig:expsetup}-c). Note that next to parallelization through streams, algorithms performed in these streams can be highly parallelized themselves, for example vector multiplications and \glspl{FFT}. A combination of these approaches is used to fully utilize \gls{GPU} resources even when sequential time-domain algorithms such as \gls{DDLMS} are used.

\begin{figure}
    \vspace{-3mm}
    \renewcommand\figurename{Fig.}
    \centering
    \begin{subfigure}{0.5\textwidth}
        \includegraphics[width=\textwidth]{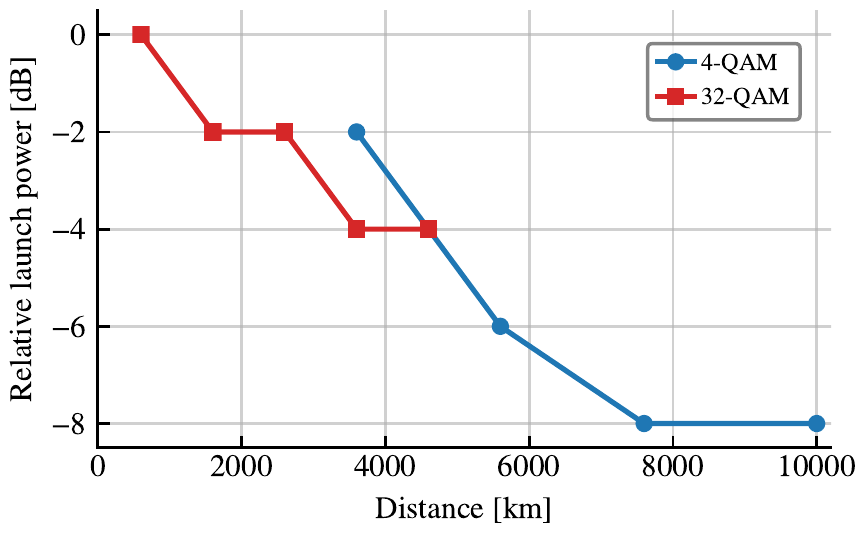}
        \vspace*{-5mm}\caption{Optimal relative launch power versus distance}
        \label{fig:optimalLP}
    \end{subfigure}%
    \begin{subfigure}{0.5\textwidth}
        \includegraphics[width=\textwidth]{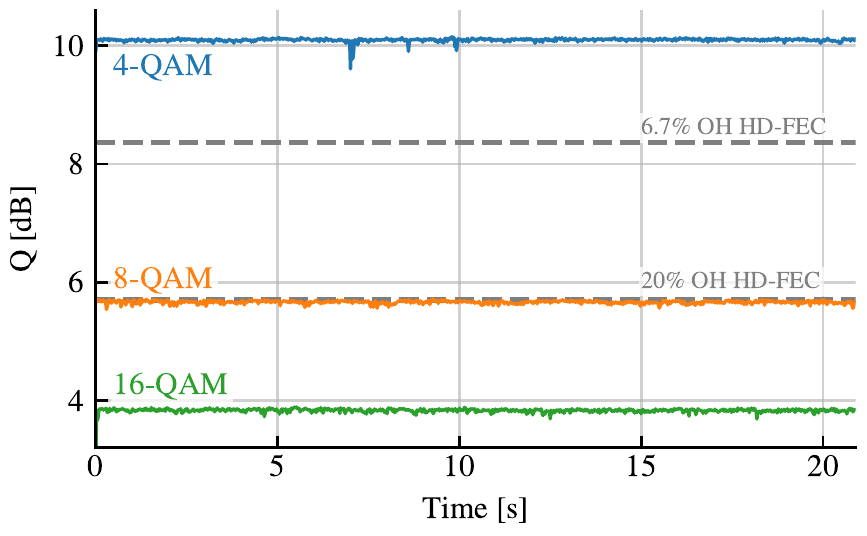}
        \vspace*{-5mm}\caption{Continuous real-time traces at \SI{7600}{km}.}
        \label{fig:longtraces}
    \end{subfigure}%
    \vspace{-5mm}
    \caption{Experimental results.}
    \label{fig:resultstop2}
    \vspace{-5mm}
\end{figure}

\section{Results and Discussion}
\cref{fig:distance} shows the Q-factor versus distance for the considered modulation formats. For each point, the launch power of the test channel with respect to the dummy channels and its \gls{CSPR} were optimized. It is shown that \qam{4} reaches the 6.7\% overhead \gls{HDFEC} threshold \cite{agrell_information-theoretic_2018} after \SI{10000}{km} of transmission. \qam{8} can be successfully transmitted over up to \SI{7600}{km} when 20\% overhead \gls{HDFEC}\cite{agrell_information-theoretic_2018} is employed. Using this threshold, \qam{16} reaches \SI{5600}{km}, \qam{32} reaches \SI{3600}{km}, and \qam{64} reaches \SI{1600}{km}. \cref{fig:launchpower} shows the dependence of the Q-factor on the relative launch power for all the considered modulation formats at their corresponding limit distances. It is shown that the performance varies substantially with the relative launch power, even at relative powers well below those of the dummy channels. This behavior suggests a dominant impact of \gls{SPM}, perhaps on the \gls{MP} carrier tone. 

\cref{fig:optimalLP} shows the evolution of the optimum launch power with transmission distance for \qam{4} and \qam{32}. In both cases, the optimum launch power decreases substantially with transmission distance. In conventional systems, the optimum launch power tends to remain independent of the transmission distance as the degradation imposed by fiber nonlinearities and noise scale at a similar rate. However, in the case of a MP signal, the nonlinear interactions between the carrier and signal components of the signal seem to scale at a substantially higher rate, leading to a decrease of the optimum launch power with transmission distance. Further investigation of the impact of fiber nonlinearities in MP signals is required to fully describe this behavior.


Finally, \cref{fig:longtraces} shows continuous real-time performance of the receiver for \qam{4-, 8-, and 16} during 20 second periods. The short-term averaged Q-factors were estimated from \gls{BER} in bins of \SI{21}{ms}. It is shown that the system was capable of sustained operation with the main limitation imposed by the RAM capacity of our workstation.

\glsreset{MP}
\section{Conclusions}
We demonstrate real-time operation of \gls{MP} \gls{KK} \qam{N} at \SI{1}{GBaud} over an experimental \SI{10000}{km} straight-line optical fiber link. \qam{4} is successfully transmitted over \SI{10000}{km}, with \qam{8} reaching \SI{7600}{km}, \qam{16} reaching \SI{5600}{km}, \qam{32} reaching \SI{3600}{km}, and \qam{64} reaching \SI{1600}{km}. We also addressed the impact of launch power on the system performance, showing that the transmission of \gls{MP} signals differs substantially from that of conventional systems. Finally, stable performance for \qam{4-, 8-, and 16} in 20 second long continuous real-time traces over \SI{7600}{km} was shown. These results demonstrate the potential enabled by massive parallel processing using \glspl{GPU} in optical communications, including long distance systems.
\\
\emph{\footnotesize{Partial funding is from the Dutch NWO Gravitation Program on Research Center for Integrated Nanophotonics (Grant Number 024.002.033). In addition, we would like to thank Mr. Hiroyuki Sumimoto and Mr. Koichi Suto for technical support with the experimental setup.}}
\vspace{-4mm}

\bibliographystyle{style/osajnl}
\bibliography{ref.bib}

\end{document}